\documentclass[aps,prl,showpacs,twocolumn]{revtex4-1}
\usepackage{amsfonts}
\usepackage{amssymb}
\usepackage{amsmath}
\usepackage{graphicx}
\usepackage{epsfig}
\usepackage{subfigure}
\usepackage{appendix}
\usepackage{color}
\usepackage{braket}
\usepackage{hyperref}
\begin{document}

\title{Exact polariton-condensate states via nonlinear Stark coupling}
\author{D. K. He}
\author{Z. Song}
\email{songtc@nankai.edu.cn}
\affiliation{School of Physics, Nankai University, Tianjin 300071, China}

\begin{abstract}
We study a hybrid system of coupled cavities, with or without two-level atoms doped inside. By employing a building-block construction facilitated by nonlinear Stark coupling, we identify a manifold of exact many-body eigenstates that remain analytically tractable across arbitrary lattice geometries and atom distributions. We demonstrate that these exact states exhibit a dual physical nature depending on their spectral positioning: when embedded within the chaotic many-body spectrum, they manifest as quantum scars that violate the eigenstate thermalization hypothesis and undergo weak ergodicity breaking; conversely, as ground states within fixed-total-excitation sectors, they exhibit off-diagonal long-range order and form polariton condensates. Furthermore, numerical calculations of local excitation-number fluctuations suggest that negative Stark coupling shifts the Mott-insulator to superfluid transition toward weaker inter-cavity hopping. Our findings provide a versatile framework for engineering both coherent quantum phases and non-thermalizing states in diverse light-matter architectures.
\end{abstract}

\maketitle

\textit{Introduction}.---Cavity quantum electrodynamics (QED) provides tunable photon--atom platforms for strongly interacting many-body physics \cite{carusotto2013quantum,raimond2001manipulating,blais2021circuit}. Their complexity makes exact eigenstates valuable for understanding correlations and benchmarking numerical methods \cite{schmidt2009strong,affleck1987rigorous,sutherland2004beautiful,shiraishi2017systematic}. Coupled-cavity arrays are commonly described by the Jaynes--Cummings (JC) framework \cite{jaynes2005comparison,angelakis2007photon,hartmann2006strongly}. Nonlinear Stark coupling extends this framework and has been explored in Dicke models \cite{grimsmo2013cavity,chen2024phase,zhai2025stark,wang2026superradiant} and topological phase transitions \cite{ying2023scaling,ying2024spin}. Although nonlinear interactions generally complicate analytical treatments, we show that Stark coupling can instead enable exact many-body constructions in hybrid photon--atom lattices.

Quantum many-body scars (QMBS) are nonthermal eigenstates embedded in otherwise thermalizing spectra, violating the eigenstate thermalization hypothesis (ETH) and producing weak ergodicity breaking \cite{turner2018weak,serbyn2021quantum,deutsch1991quantum,srednicki1994chaos,d2016quantum}. Experimental observations \cite{bernien2017probing,bluvstein2021controlling} motivate analytical constructions based on spectrum generating algebra (SGA), restricted spectrum generating algebra (RSGA) \cite{moudgalya2020eta,moudgalya2018exact,schecter2019weak,mark2020eta,liu2025condensate}, and projection operators \cite{shiraishi2017systematic,dooley2025perfect}. Hilbert-space fragmentation provides another route to nonergodic dynamics \cite{khemani2020localization,sala2020ergodicity,wang2025exploring,he2026hilbert}. These approaches have primarily addressed spin and single-species models \cite{moudgalya2022quantum}, leaving exact scar constructions in photon--atom systems comparatively unexplored.

Here we extend the building-block method \cite{liu2026exceptional} to photon--atom lattices with unbounded local photonic Hilbert spaces. Nonlinear Stark coupling enables an RSGA-protected tower of exact states for arbitrary lattice geometries and atom distributions under compatible resonance conditions, without requiring translation or reflection symmetry. On the ground-state branch, these states exhibit off-diagonal long-range order (ODLRO) \cite{penrose1956bose,yang1962concept,zhang2025exact,zhang2022steady,he2025eta} within fixed-excitation sectors; embedded in a chaotic spectrum, they form a low-entanglement scar tower. We further use local excitation-number fluctuations to investigate the effect of Stark coupling on the Mott-insulator (MI)--superfluid (SF) transition.

\textit{Model and Hamiltonian construction}.---We consider $N$ coupled single-mode cavities arranged on an arbitrary lattice, each either containing a two-level atom or remaining empty. In an atom-doped cavity, the photon mode interacts with the atom through both JC coupling and a nonlinear Stark interaction, while photons hop between adjacent cavities. In the absence of Stark coupling, a uniformly atom-doped array with uniform physical onsite parameters recovers the JC-Hubbard (JCH) model \cite{greentree2006quantum}, whose MI--SF transition has been extensively studied \cite{angelakis2007photon,hartmann2006strongly}. Below, we exploit nonlinear Stark coupling to construct lattice Hamiltonians with exact many-body eigenstates using the building-block method.

\begin{figure}[bh]
\centering
\includegraphics[width=0.46\textwidth]{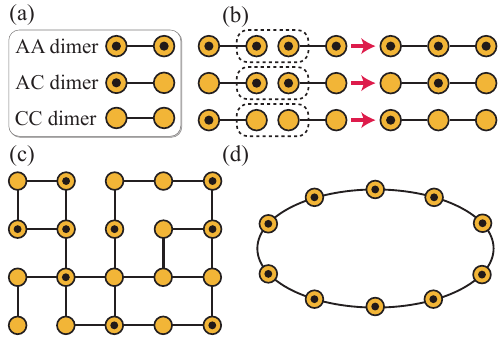}
\caption{Construction of the hybrid system Hamiltonian via the building-block method. (a) The three fundamental dimer units used for Hamiltonian construction: the AA dimer, AC dimer, and CC dimer, as defined in Eq. (\ref{3_dimers}). (b) Illustrative examples of Hamiltonian composition through dimer overlapping. The connection between identical or distinct dimers is realized by merging the sites enclosed in the dashed boxes. Consistency requires that the overlapping sites be of the same species (either both atom-doped or both empty). (c) Schematic representation of an arbitrary lattice system formed by a sequence of arbitrary dimers. (d) Schematic representation of a one-dimensional periodic chain assembled from a sequence of identical AA dimers.}
\label{fig1}
\end{figure}

The resonant Hamiltonian is assembled from overlapping dimers,
\begin{equation}
H_{\text{res}}=\sum_{\langle i,j\rangle}H_{ij}.
\end{equation}
The building blocks comprise three types: AA dimers, in which both cavities contain an atom; AC dimers, in which only one cavity is atom-doped; and CC dimers, in which both cavities are empty [Fig.~\ref{fig1}(a)]. Their Hamiltonians are
\begin{equation}
H_{ij}=
\begin{cases}
h_i^{\mathrm{a}}+h_j^{\mathrm{a}}
-\kappa(a_i^\dagger a_j+a_j^\dagger a_i),
& \mathrm{AA},\\[1mm]
h_i^{\mathrm{a}}+h_j^{\mathrm{c}}
-\kappa(a_i^\dagger a_j+a_j^\dagger a_i),
& \mathrm{AC},\\[1mm]
h_i^{\mathrm{c}}+h_j^{\mathrm{c}}
-\kappa(a_i^\dagger a_j+a_j^\dagger a_i),
& \mathrm{CC},
\end{cases}
\label{3_dimers}
\end{equation}
where the two local contributions are defined as
\begin{equation*}
\begin{aligned}
h_l^{\mathrm{a}}={}&
\omega_l a_l^\dagger a_l+\Omega_l s_l^z\\
&+g_l(a_l^\dagger s_l^-+a_l s_l^+)
+U_l s_l^z a_l^\dagger a_l,\\
h_l^{\mathrm{c}}={}&
\varpi_l a_l^\dagger a_l.
\end{aligned}
\end{equation*}
Here, $a_l^\dagger$ ($a_l$) creates (annihilates) a photon in cavity $l$, whose photonic vacuum is denoted by $|0\rangle_l$. The pseudospin operators $s_l^\alpha$ describe the atomic degrees of freedom, with $s_l^z|e\rangle_l=\frac{1}{2}|e\rangle_l$ and $s_l^z|g\rangle_l=-\frac{1}{2}|g\rangle_l$. The parameters entering $h_l^{\mathrm{a,c}}$ specify the local contributions assigned to an individual dimer; in particular, $g_l$ and $U_l$ characterize the JC and nonlinear Stark couplings within that block. We take the photon hopping amplitude to be $\kappa>0$.

Identical or distinct dimers can be connected by merging the sites enclosed in the dashed boxes in Fig.~\ref{fig1}(b). A shared site must have the same cavity type and the same local term in every incident dimer. These local contributions add upon assembly: a site shared by $d_l$ dimers contributes $d_l h_l^{\mathrm{a,c}}$ to the full Hamiltonian. Thus, the physical onsite coefficients of the assembled lattice are obtained by summing the contributions from all incident dimers.

To preserve a common family of exact states under this assembly, we impose compatible algebraic conditions on the building blocks. Choosing the exact-state ladder spacing of each dimer to vanish yields the sufficient resonance conditions
\begin{equation}
\begin{aligned}
g_l&=\pm\sqrt{2\Omega_l(\omega_l-\kappa)},\\
U_l&=-2(\omega_l-\kappa),\qquad
\varpi_l=\kappa.
\end{aligned}
\label{Condition_0}
\end{equation}
The first two conditions apply to atom-doped sites, and the third applies to empty cavities. Under these conditions, the three types of dimers can be assembled into lattices with different geometries and atom distributions, as exemplified in Fig.~\ref{fig1}(c). A periodic chain is obtained by joining identical AA dimers into a ring, with every site shared by two dimers [Fig.~\ref{fig1}(d)].

The resonance conditions imply $g_l^2=-\Omega_l U_l$. For real, nonzero $g_l$, the negative-$U_l$ branch requires $\Omega_l>0$, whereas the positive-$U_l$ branch requires $\Omega_l<0$. When $U_l<0$ at every atom-doped site, the constructed exact states are ground states within their respective fixed-total-excitation sectors. The excited-state properties of the positive-$U_l$ branch are examined below. Within this resonant construction, setting $U_l=0$ also forces $g_l=0$, decoupling light and matter. Nonzero Stark coupling is therefore essential for the nontrivial hybrid exact states obtained by this construction. The RSGA relations, the compatibility conditions underlying the building-block method, and a proof of the ground-state property within each fixed-total-excitation sector are provided in the Supplemental Material (SM) \cite{SM}.

\begin{figure*}[th]
\centering
\includegraphics[width=0.9\textwidth]{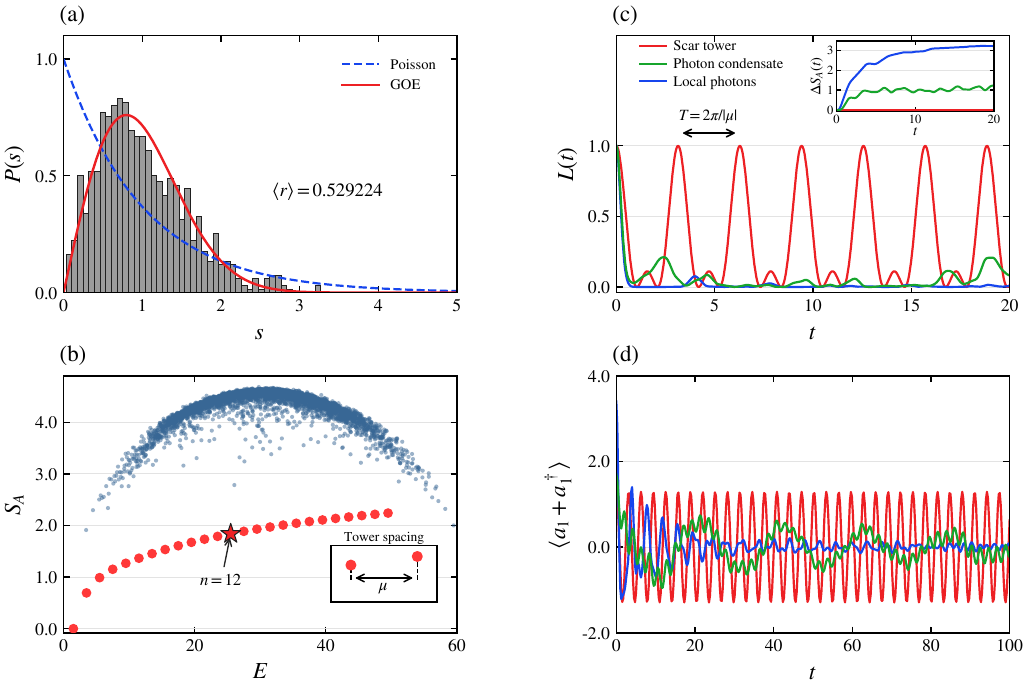}
\caption{
Spectral statistics and dynamics of the exact scar tower.
(a) Level-spacing distribution $P(s)$ in the middle of the
spectrum in the even-reflection-parity sector with
$\mathcal{N}=12$, compared with GOE (red solid) and Poisson
(blue dashed) predictions; $\langle r\rangle=0.529224$.
(b) Bipartite entanglement entropy $S_A$ versus energy $E$.
Blue points represent eigenstates in the $\mathcal{N}=12$
sector, while red circles show the exact-state tower for
$n=0,\ldots,24$; the star identifies its $n=12$ member.
Inset: adjacent tower states separated by $\mu$.
(c) Loschmidt echo $L(t)$, with scar revival period
$T=2\pi/\mu=\pi$.
Inset: entanglement-entropy change
$\Delta S_A(t)=S_A(t)-S_A(0)$.
(d) Local field quadrature $\langle a_1+a_1^\dagger\rangle$.
In (c), its inset, and (d), the red, green, and blue curves
correspond, respectively, to the initial states
$|\Psi_{\mathrm{sc}}\rangle=
3^{-1/2}\sum_{n=5}^{7}|\psi_{\mathrm{E}}^n\rangle$,
$|\Psi_{\mathrm{ph}}\rangle=
3^{-1/2}\sum_{n=5}^{7}
\frac{(\sum_{j=1}^{N}a_j^\dagger)^n}{\sqrt{N^n n!}}
|\mathrm{Vac}\rangle$,
and
$|\Psi_{\mathrm{loc}}\rangle=
3^{-1/2}\sum_{n=5}^{7}
\frac{(a_1^\dagger)^n}{\sqrt{n!}}|\mathrm{Vac}\rangle$.
Panels (a)--(d) use an open chain with $N=4$ and an equal
bipartition $N_A=N_B=2$ for the entanglement entropy.
The Hamiltonian parameters are
$\omega=-\Omega=1$, $g=\sqrt{2}$, $\kappa=1$,
$U=2$, and $\mu=2$.
All three initial states evolve under the same
$H_{\mathrm{chain}}$.}
\label{fig2}
\end{figure*}
Under the aforementioned resonance conditions, the exact many-body states of the system are given by
\begin{equation}
|\psi_{\text{E}}^{n}\rangle = \frac{1}{\sqrt{\Lambda_{n}}} \eta^{n} |\text{Vac}\rangle,
\end{equation}
where $\Lambda_{n}$ is the normalization factor and $|\text{Vac}\rangle = \prod_{l=1}^{N} |g,0\rangle_{l}$ represents the collective vacuum state. These states satisfy $H_{\text{res}}\ket{\psi_{\text{E}}^{n}}=E_{\mathrm{vac}}\ket{\psi_{\text{E}}^{n}}$ for any non-negative integer $n$, where
$E_{\mathrm{vac}}\equiv\langle\text{Vac}|H_{\text{res}}|\text{Vac}\rangle=-\frac{1}{2}\sum_{l=1}^{N}d_l\Omega_l$
is the vacuum energy. Here, $d_l$ denotes the number of dimers sharing the atom-doped site $l$, and hence the number of times its local term is counted in $H_{\text{res}}$. For empty cavities, $\Omega_l$ is understood to be zero in this sum. The collective creation operator $\eta$ is defined as the sum of local operators, $\eta = \sum_{l=1}^{N} \eta_{l}$, where
\begin{equation}
\eta_{l} = \begin{cases}
a_{l}^{\dagger} + \frac{U_{l}}{g_{l}}s_{l}^{+}, & \text{for cavities with atoms (} h_{l}^{\text{a}} \text{)} \\
a_{l}^{\dagger}, & \text{for empty cavities (} h_{l}^{\text{c}} \text{)}
\end{cases}.
\end{equation}
Since the Hamiltonian commutes with the total excitation operator, $[H_{\text{res}}, \hat{\mathcal{N}}] = 0$, where $\hat{\mathcal{N}} = \sum_{j=1}^{N} (a_{j}^{\dagger}a_{j} + s_{j}^{+}s_{j}^{-})$, the total excitation number $n$ is a conserved quantum number. In each invariant subspace with $\mathcal{N}=n$, $|\psi_{\text{E}}^{n}\rangle$ constitutes the exact state. By introducing a chemical potential term, $H = H_{\text{res}} + \mu \hat{\mathcal{N}}$, these states form an equidistant energy tower with eigenvalues $E_n = E_{\mathrm{vac}}+n\mu$. The RSGA preserves this exact tower even in the absence of translation or reflection symmetry, provided that the building-block resonance conditions are maintained. Its quantum-scar properties are examined below.

\textit{Quantum Scar Tower and Off-Diagonal Long-Range Order Ground State}.---The exact states constructed via the building-block method possess unique spectral and coherence properties that distinguish them from typical many-body states.

\textbf{Quantum scar tower}.---A distinctive feature of the exact eigenstates $|\psi_{\text{E}}^{n}\rangle$ is their identification as many-body quantum scars embedded within a chaotic spectrum. To investigate this, we consider an open boundary chain constructed by concatenating identical AA dimers
\begin{eqnarray}
H_{\textrm{chain}}&=&\sum_{j=1}^{N-1}H_{j,j+1}^{\mathrm{AA}}+ \mu\hat{\mathcal{N}}, \\
h_{j}^{\mathrm{a}}&=&\frac{\omega}{2}a_{j}^{\dagger }a_{j}+\frac{\Omega}{2}s_{j}^{z}+\frac{g}{2}( a_{j}^{\dagger }s_{j}^{-}+a_{j}s_{j}^{+}) + \frac{U}{2}s_{j}^{z}a_{j}^{\dagger }a_{j}. \notag
\end{eqnarray}
Open boundaries remove translation symmetry. Within a fixed excitation and even-reflection-parity sector, the normalized spacings $s_n=(E_{n+1}-E_n)/\langle\Delta E\rangle$ agree with the Gaussian Orthogonal Ensemble (GOE) prediction $P(s)=\frac{\pi}{2}s\exp(-\frac{\pi}{4}s^2)$ [Fig.~\ref{fig2}(a)], supporting a chaotic spectral background \cite{li2021eigenstate,ma2022ergodic}. In contrast, the exact states have anomalously low bipartite entropy $S_A=-\mathrm{Tr}(\rho_A\ln\rho_A)$ relative to surrounding eigenstates [Fig.~\ref{fig2}(b)], and form a scar tower with spacing $\mu$.

\begin{figure}[]
\centering
\includegraphics[width=0.46\textwidth]{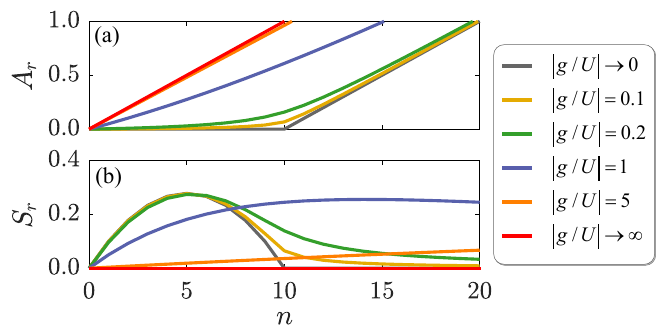}
\caption{Photon and atomic correlations of the exact
ground state $\ket{\psi_{\mathrm G}^{n}}$ for $N=10$
at $U=U_R$.
Panels (a) and (b) show $A_r$ and $S_r$, respectively,
as functions of the total excitation number $n$.
Different colors denote different values of $|g/U|$.
Both correlators are independent of the separation
between distinct sites.}
\label{fig3}
\end{figure}

\begin{figure*}[th]
\centering
\includegraphics[width=1\textwidth]{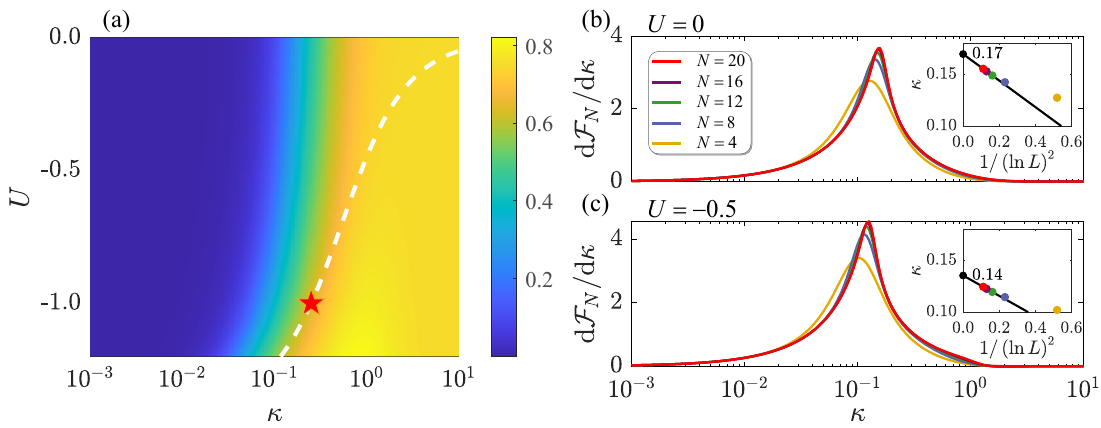}
\caption{
Ground-state excitation-number fluctuations and finite-size estimates of the MI--SF transition for $H_{\mathrm{ring}}$ at unit filling, $\mathcal{N}=N$. (a) Site-averaged variance $\mathcal{F}_N$, defined in Eq.~(\ref{Fluctuation}), in the $\kappa$--$U$ plane for $N=4$. The white dashed curve denotes the resonance condition $U=U_R$. The red star at $(\kappa,U)=(1/4,-1)$ marks the point where the exact ground-state energies are degenerate across excitation sectors. (b),(c) Derivatives $\partial\mathcal{F}_N/\partial\kappa$ obtained using DMRG for $U=0$ and $U=-0.5$, respectively, with $N=4,8,12,16$, and $20$. The insets show the peak positions versus $1/(\ln L)^2$, with $L=N$. Linear fits to the three largest sizes yield fluctuation-based transition estimates of $\kappa_{\mathrm{est}}\simeq0.17$ and $0.14$, respectively. Throughout, $\Omega=\omega=g=1$.}
\label{fig4}
\end{figure*}

We characterize the dynamics of the three initial states through the Loschmidt echo and local field coherence. For an initial state $|\Psi(0)\rangle$, the time-evolved state is $|\Psi(t)\rangle=e^{-iH_{\mathrm{chain}}t}|\Psi(0)\rangle$,
and the Loschmidt echo is defined as $ L(t)=|\langle\Psi(0)|\Psi(t)\rangle|^2$. The scar superposition $|\Psi_{\mathrm{sc}}\rangle$ exhibits perfect revivals with period $T=2\pi/\mu$ [Fig.~\ref{fig2}(c)], constant entanglement entropy [inset of Fig.~\ref{fig2}(c)], and persistent oscillations of the local field quadrature $\langle a_1+a_1^\dagger\rangle$ [Fig.~\ref{fig2}(d)]. The photon condensate $|\Psi_{\mathrm{ph}}\rangle$, which has a larger overlap with the scar tower than the localized photon state, exhibits partial revivals and retains memory of its initial local coherence over longer times. In contrast, the localized photon state $|\Psi_{\mathrm{loc}}\rangle$, with a much smaller scar-tower overlap, displays a rapid decay of the return probability and local coherence, accompanied by pronounced entanglement growth. This rapid relaxation is consistent with thermalizing behavior in the chaotic spectral background.

\textbf{Off-diagonal long-range order ground state}.---At finite excitation density, the constructed exact ground states exhibit ODLRO. To characterize their coherence explicitly, we consider a uniformly atom-doped cavity ring assembled from identical AA dimers overlapping end-to-end, as illustrated in Fig.~\ref{fig1}(d). The corresponding Hamiltonian is given by
\begin{eqnarray}
H_{\textrm{ring}}&=&\sum_{j=1}^{N}H_{j,j+1}^{\mathrm{AA}},\label{Hring}
\end{eqnarray}
with periodic boundary conditions (PBC) $a_{N+1}=a_{1}$.
Under the PBC resonance conditions $U=U_{R}$, where $U_{R}=\Omega-\omega+2\kappa-\sqrt{\left(\Omega_{}-
\omega+2\kappa\right)^2+2g^2}$, a set of ground states
in each invariant subspace can be explicitly written as%
\begin{equation}
\left\vert \psi _{\textrm{G}}^{n}\right\rangle =\frac{1}{\sqrt{%
\Lambda _{n}}}\left[ \sum_{j=1}^{N}\left( ga_{j}^{\dag }+U s_{j}^{+}\right) \right] ^{n}\left\vert \mathrm{Vac}\right\rangle,
\end{equation}
where the PBC resonance conditions and the normalization factor are given in SM \cite{SM}.

We characterize the coherence of $\ket{\psi_{\mathrm G}^{n}}$ by the photon correlation $A_r=\langle a_j^\dagger a_{j+r}\rangle$ and the atomic correlation $S_r=\langle s_j^+s_{j+r}^-\rangle$, where
the expectation values are taken in $\ket{\psi_{\mathrm G}^{n}}$ and $r=1,\ldots,N-1$, with site indices understood modulo $N$. In the limit $U_R/g\to0$, the state contains only
photonic excitations, yielding $A_r=n/N$ and $S_r=0$.
Conversely, in the limit $g/U_R\to0$ with
$0\leq n\leq N$, it reduces to a symmetric atomic state
with $n$ excitations, for which $A_r=0$ and
$S_r=n(N-n)/[N(N-1)]$.

For nonzero $g$ and $U_R$, both correlation functions
are independent of the separation $r$.
More importantly, at fixed positive filling
$\rho=n/N$ and fixed nonzero couplings, we derive
strictly positive lower bounds for both correlators
that are independent of the system size
(see the SM \cite{SM}).
Thus, both correlations remain finite and nonzero
at arbitrarily large separations in the thermodynamic
limit, establishing simultaneous photon and atomic
ODLRO in the exact ground state. The dependence of the photon and atomic correlations
on the total excitation number $n$ is shown in
Figs.~\ref{fig3}(a) and \ref{fig3}(b), respectively,
for $N=10$ and different values of $|g/U_R|$.

\textit{Mott-insulator to superfluid transition}.--- We investigate the MI--SF transition of $H_{\mathrm{ring}}$ at unit filling, $\mathcal{N}=N$, using local excitation-number fluctuations, following earlier studies of coupled JC cavities \cite{angelakis2007photon}. We define the site-averaged variance
\begin{equation}
\mathcal{F}_{N}
=\frac{1}{N}\sum_{j=1}^{N}
\left[
\langle \hat{\mathcal{N}}_{j}^{2}\rangle
-\langle \hat{\mathcal{N}}_{j}\rangle^{2}
\right],
\label{Fluctuation}
\end{equation}
where $\hat{\mathcal{N}}_{j}=a_j^\dagger a_j+s_j^+s_j^-$ and the expectation values are evaluated in the ground state within the unit-filling sector. Local fluctuations are suppressed deep in the MI regime and increase as excitations become more delocalized.

The numerical results for $N=4$ illustrate this redistribution in the $\kappa$--$U$ plane [Fig.~\ref{fig4}(a)]. The resonance line $U=U_R$, indicated by the white dashed curve, provides an analytical reference along which the exact ground states possess ODLRO. Away from this line, the fluctuation map shows that negative Stark coupling enhances local fluctuations and shifts their rapid increase toward weaker hopping.

To estimate the transition location, we calculate $\partial\mathcal{F}_{N}/\partial\kappa$ using DMRG
for $N=4,8,12,16$, and $20$ [Figs.~\ref{fig4}(b) and \ref{fig4}(c)]. Taking the peak positions $\kappa_p(N)$ as finite-size proxies for the transition, we apply the BKT-motivated extrapolation $\kappa_p(N)=\kappa_{\mathrm{est}}+A/(\ln N)^2$ to the three largest sizes. The resulting fluctuation-based estimates are $\kappa_{\mathrm{est}}\simeq0.17$ for $U=0$ and $\kappa_{\mathrm{est}}\simeq0.14$ for $U=-0.5$, as shown in the insets. These results are consistent with a Stark-induced shift of the MI--SF transition toward weaker inter-cavity hopping.

\textit{Conclusion}.--- We have developed an RSGA-based building-block construction of exact many-body states in photon--atom systems with nonlinear Stark coupling, applicable to arbitrary lattice geometries and atom distributions. These states realize quantum scars with coherent revivals in a chaotic spectral background and polariton condensates with ODLRO on the ground-state branch. Our fluctuation-based numerical analysis further suggests that negative Stark coupling shifts the MI--SF transition toward weaker inter-cavity hopping.

\textit{Acknowledgments}---This work was supported by the National Natural Science Foundation of China (under Grant No. 12374461).

\textit{Data availability}---The data that support the findings of this article are openly available \cite{He2026Data}.

\bibliographystyle{unsrt}
\bibliography{ref.bib}

\end{document}


\begin{center}
{\large \textbf{Supplemental Material for
``Exact polariton-condensate states via nonlinear Stark coupling'' }}

D. K. He and Z. Song*\\[2pt]
\textit{School of Physics, Nankai University, Tianjin 300071, China}\\[0pt]
*songtc@nankai.edu.cn
\end{center}

\setcounter{equation}{0} \renewcommand{\theequation}{S\arabic{equation}} %
\setcounter{figure}{0} \renewcommand{\thefigure}{S\arabic{figure}} %
\setcounter{secnumdepth}{3}

In this Supplemental Material, we present:
A. the derivation of the restricted spectrum generating algebra
(RSGA) in the single-cavity Jaynes--Cummings (JC) model with
nonlinear Stark coupling;
B. the building-block construction of exact eigenstates for
arbitrary lattice geometries and atom distributions, together
with its extension to periodic rings;
C. a proof that the constructed exact states are ground states
within their respective fixed-total-excitation sectors under
the specified resonance conditions and on the negative-Stark-coupling
branch;
D. analytical derivations of the normalization factor and the
photon and atomic correlation functions used to characterize
off-diagonal long-range order (ODLRO).

\subsection{RSGA in the JC model with nonlinear Stark interactions}

\label{A} We consider a JC model with nonlinear Stark interactions in a
single cavity, whose Hamiltonian is given by : 
\begin{eqnarray}
h_{l}^{\text{a}} = \omega _{l}a_{l}^{\dagger }a_{l}+\Omega
_{l}s_{l}^{z}+g_{l}\left( a_{l}^{\dagger
}s_{l}^{-}+a_{l}s_{l}^{+}\right)+U_{l}s_{l}^{z}a_{l}^{\dagger }a_{l}.
\end{eqnarray}
We can verify that for the operator 
\begin{eqnarray}
\eta_{l}=g_{l}a_{l}^{\dagger}+U_{l}s_{l}^{+},
\end{eqnarray}
the Hamiltonian possesses a first-order RSGA when $U_{l}=\left(\Omega_{l}-
\omega_{l}\pm\sqrt{\left(\Omega_{l}-\omega_{l}\right)^2+2g_{l}^2} \right)$.
The proof is as follows. We examine the first-order commutator 
\begin{eqnarray}
\left[h_{l}^{\text{a}},\eta_{l}\right]=g_{l}\omega_{l}a_{l}^{
\dagger}-g_{l}U_{l}s_{l}^{z}a_{l}^{\dagger}+\left(g_{l}^{2}+\Omega_{l}U_{l}%
\right)s_{l}^{+}+U_{l}^{2}s_{l}^{+}a_{l}^{\dagger}a_{l},  \label{Single_Com}
\end{eqnarray}
whose action on the state$\ket{g,0}_{l}$ yields 
\begin{eqnarray}
\left[h_{l}^{\text{a}},\eta_{l}\right]\ket{g,0}_{l}=\left(g_{l}\omega_{l}+%
\frac{1}{2}g_{l}U_{l}\right) \ket{g,1}_{l}+\left(g_{l}^{2}+\Omega_{l}U_{l}%
\right)\ket{e,0}_{l}.
\end{eqnarray}
We have calculated the second-order commutator 
\begin{equation}
\left[\left[h_{l}^{\text{a}},\eta_{l}\right],\eta_{l}\right]=0.
\end{equation}
To satisfy the RSGA condition, it must hold that 
\begin{eqnarray}
\left[h_{l}^{\text{a}},\eta_{l}\right]\ket{g,0}_{l}=\epsilon_{l}\eta_{l}%
\ket{g,0}_{l}.
\end{eqnarray}
This leads to the following constraint equation on the parameters: 
\begin{equation}
\left(g_{l}^{2}+\Omega_{l}U_{l}\right)=U_{l}\left(\omega_{l}+\frac{1}{2}%
U_{l}\right).
\end{equation}
When $U_{l}=\left(\Omega_{l}-\omega_{l}\pm\sqrt{\left(\Omega_{l}-\omega_{l}
\right)^2+2g_{l}^2}\right)$ the Hamiltonian satisfies the RSGA condition on
the state $\ket{g,0}_{l}$. The energy difference of the states protected by
the RSGA algebra is $\epsilon_{l}=\omega_{l}+\frac{1}{2}U_{l}$. The
corresponding state takes the form 
\begin{equation}
\ket{\psi_n}=\frac{1}{\sqrt{\Lambda}}\left(\eta_{l}\right)^n \ket{g,0}_{l}.
\end{equation}
By observing the relation satisfied by $U_{l}$, we can find that $U_{l}$ has two branches of solutions (positive and negative). The required Stark coupling is determined jointly by the JC coupling and the detuning. When $U_l$ takes the negative branch, the RSGA-protected state is a ground state of $h_l^{\mathrm a}$ within the corresponding fixed-total-excitation sector.

\subsection{Three types of dimers and the building-block method}

\label{B} We first prove that for the three types of dimers-the basic units of the building-block method-each has its own
RSGA (SGA) condition. In the building-block construction below, the local generators are understood as $\eta_l=a_l^\dagger+(U_l/g_l)s_l^+$ at atom-doped sites and $\eta_l=a_l^\dagger$ at empty sites. These three fundamental dimers are constructed from two types of basic cavity units: the aforementioned $h_{l}^{a}$ and an empty cavity 
\begin{equation}
	h_{l}^{c} = \varpi_{l} a_{l}^{\dagger} a_{l}.
\end{equation}
We shall prove this in sequence.

(i) For the AA dimer: 
\begin{equation}
H_{ij}^{\mathrm{AA}}=h_{i}^{\mathrm{a}}+h_{j}^{\mathrm{a}}-\kappa
\left(a_{i}^{\dagger}a_{j}+a_{j}^{\dagger}a_{i}\right).
\end{equation}
For the operator $\eta_{ij}=a_{i}^{\dagger}+\frac{U_{i}}{g_{i}}%
s_{i}^{+}+a_{j}^{\dagger}+\frac{U_{j}}{g_{j}}s_{j}^{+}$, we can verify the
first- and second-order commutators 
\begin{eqnarray}
\left[H^{\mathrm{AA}}_{ij},\eta_{ij}\right]&=&\left(\omega_{i}-\kappa\right)a_{i}^{
\dagger}-U_{i}s_{i}^{z}a_{i}^{\dagger}+\left(g_{i}+\Omega_{i}\frac{U_{i}}{%
g_{i}} \right)s_{i}^{+}+\frac{U_{i}^{2}}{g_{i}}s_{i}^{+}a_{i}^{\dagger}a_{i}
\notag \\
&+&\left(\omega_{j}-\kappa\right)a_{j}^{
\dagger}-U_{j}s_{j}^{z}a_{j}^{\dagger}+\left(g_{j}+\Omega_{j}\frac{U_{j}}{%
g_{j}} \right)s_{j}^{+}+\frac{U_{j}^{2}}{g_{j}}s_{j}^{+}a_{j}^{\dagger}a_{j},
\notag \\
\left[\left[H^{\mathrm{AA}}_{ij},\eta_{ij}\right],\eta_{ij}\right]&=&0.
\label{AA_Condition}
\end{eqnarray}
We find that the result differs little from Eq. (\ref{Single_Com}), except
that the corresponding $\omega_{l}$ is replaced by $\omega_{l}-\kappa$. This
indicates that the coupling between different cavities affects the effective
chemical potential, which in turn influences the strengths of the Stark
interactions $U_{i}$ and $U_{j}$ required to satisfy the RSGA. The strength
of the Stark interaction for each cavity is: $U_{i(j)}=\left(\Omega_{i(j)}-%
\omega_{i(j)}+\kappa\pm\sqrt{\left(\Omega_{i(j)}-\omega_{i(j)}+\kappa%
\right)^2+2g_{i(j)}^2}\right)$. Meanwhile, the condition $\omega_{i}+\frac{%
U_{i}}{2}=\omega_{j}+\frac{U_{j}}{2}$ must also be satisfied. The energy
level splitting between RSGA-protected states in invariant subspaces with
different excitation numbers is $\omega_{i}-\kappa+\frac{U_{i}}{2}$.
Similarly, when $U_{i(j)}$ takes the negative solution, the RSGA-protected
state becomes the ground state in the subspace with fixed excitation number. When $U_{l}$ takes the positive branch solution, the RSGA-protected state is a highly excited state of the Hamiltonian.

(ii) For the AC dimer: 
\begin{equation}
H_{ij}^{\mathrm{AC}}=h_{i}^{\mathrm{a}}+h_{j}^{\mathrm{c}}-\kappa
\left(a_{i}^{\dagger}a_{j}+a_{j}^{\dagger}a_{i}\right).
\end{equation}
For the operator $\eta_{ij}=a_{i}^{\dagger}+\frac{U_{i}}{g_{i}}%
s_{i}^{+}+a_{j}^{\dagger}$, we can verify the first- and second-order
commutators 
\begin{eqnarray}
\left[H^{\mathrm{AC}}_{ij},\eta_{ij}\right]&=&\left(\omega_{i}-\kappa\right)a_{i}^{
\dagger}-U_{i}s_{i}^{z}a_{i}^{\dagger}+\left(g_{i}+\Omega_{i}\frac{U_{i}}{%
g_{i}} \right)s_{i}^{+}+\frac{U_{i}^{2}}{g_{i}}s_{i}^{+}a_{i}^{\dagger}a_{i}
\notag \\
&+&\left(\varpi_{j}-\kappa\right)a_{j}^{\dagger},  \notag \\
\left[\left[H^{\mathrm{AC}}_{ij},\eta_{ij}\right],\eta_{ij}\right]&=&0.
\end{eqnarray}
It can be found that the condition for RSGA is: $U_{i}=\left(\Omega_{i}-%
\omega_{i}+\kappa\pm\sqrt{\left(\Omega_{i}-\omega_{i}+\kappa%
\right)^2+2g_{i}^2}\right)$ and $\omega_{i}-\kappa+\frac{U_{i}}{2}%
=\varpi_{j}-\kappa$ is satisfied. The energy level splitting between
RSGA-protected states in invariant subspaces with different excitation
numbers is $\varpi_{j}-\kappa$. When $U_i$ takes the negative branch, the RSGA-protected state is likewise a ground state of $H_{ij}^{\mathrm{AC}}$ within the corresponding fixed-total-excitation sector. When $U_{l}$ takes the positive branch solution, the RSGA-protected state is likewise a highly excited state of the Hamiltonian.

(iii) For CC dimer: 
\begin{equation}
H_{ij}^{\mathrm{CC}}=h_{i}^{\mathrm{c}}+h_{j}^{\mathrm{c}}-\kappa
\left(a_{i}^{\dagger}a_{j}+a_{j}^{\dagger}a_{i}\right).
\end{equation}
For this Hamiltonian, we can verify the first- and second-order commutators
with respect to the operator $\eta_{ij}=a_{i}^{\dagger}+a_{j}^{\dagger}$ 
\begin{eqnarray}
\left[H^{\mathrm{CC}}_{ij},\eta_{ij}\right]&=&\left(\varpi_{i}-\kappa\right)a_{i}^{
\dagger}+\left(\varpi_{j}-\kappa\right)a_{j}^{\dagger},  \notag \\
\left[\left[H^{\mathrm{CC}}_{ij},\eta_{ij}\right],\eta_{ij}\right]&=&0.
\end{eqnarray}
The parameters satisfying the SGA condition are $\varpi_{i}=\varpi_{j}$ ,
with the corresponding energy level splitting being $\varpi_{i}-\kappa$.

Using these three dimers, we can construct systems with arbitrary lattice
structures. Here, we present an example of constructing a trimer using two
dimers. Here we have two dimers, $H_{ij}$ and $H_{jk}$, which require the
same species of cavity at site $j$ . They satisfy the RSGA condition with
respect to operators $\eta_{ij}=\eta_{i}+\eta_{j}$ and $\eta_{jk}=\eta_{j}+%
\eta_{k}$, respectively, i.e., 
\begin{eqnarray}
\left[H_{ij},\eta_{ij}\right]\ket{g,0}_{i}\ket{g,0}_{j}&=&\epsilon_{ij}(%
\eta_{i}+\eta_{j})\ket{g,0}_{i}\ket{g,0}_{j},  \notag \\
\left[H_{jk},\eta_{jk}\right]\ket{g,0}_{j}\ket{g,0}_{k}&=&\epsilon_{jk}(%
\eta_{j}+\eta_{k})\ket{g,0}_{j}\ket{g,0}_{k},  \notag \\
\left[\left[H_{ij},\eta_{ij}\right],\eta_{ij}\right]&=&0,  \notag \\
\left[\left[H_{jk},\eta_{jk}\right],\eta_{jk}\right]&=&0.
\end{eqnarray}
Then for the Hamiltonian $H_{ij}+H_{jk}$ with respect to the operator $%
\eta_{i}+\eta_{j}+\eta_{k}$, we have: 
\begin{eqnarray}
\left[H_{ij}+H_{jk},\eta_{i}+\eta_{j}+\eta_{k}\right]\ket{g,0}_{i}\ket{g,0}%
_{j}\ket{g,0}_{k}&=&\left[\epsilon_{ij}\eta_{i}+\left(\epsilon_{ij}+%
\epsilon_{jk}\right)\eta_{j}+\epsilon_{jk}\eta_{k}\right]\ket{g,0}_{i}%
\ket{g,0}_{j}\ket{g,0}_{k},  \notag \\
\left[\left[H_{ij}+H_{jk},\eta_{i}+\eta_{j}+\eta_{k}\right]%
,\eta_{i}+\eta_{j}+\eta_{k}\right]&=&0.  \label{BBM}
\end{eqnarray}

In particular, if $\epsilon_{ij}=\epsilon_{jk}=0$, the trimer possesses RSGA
with respect to the operator $\eta_{i}+\eta_{j}+\eta_{k}$. The corresponding
conditions are: 
\begin{eqnarray}
g_{l}&=&\pm\sqrt{2\Omega_{l}\left(\omega_{l}-\kappa\right)},  \notag \\
U_{l}&=&-2\left(\omega_{l}-\kappa\right),  \notag \\
\varpi_{l}&=&\kappa.  \label{condition}
\end{eqnarray}
This is precisely Eq. (3) in the main text. By splicing these three dimers,
one can construct Hamiltonians with arbitrary lattice structures. As long as
the cavities at different sites satisfy the condition specified in Eq. (\ref%
{condition}), the state of the system is protected by RSGA, and the
exact state is given by Eqs. (4)-(5) in the main text.

In addition to the above method for constructing RSGA, the building-block
method also provides us with a new perspective. Observing Eq. (\ref{BBM}),
cavity $j$ is counted twice. Therefore, in the previous approach we need to
set $\epsilon_{ij}=\epsilon_{jk}=0$ to satisfy the RSGA condition. A new
perspective is that we may alternatively connect all dimers end-to-end to
form a ring when building the block, such that all cavities are counted
twice. In this way, as long as all dimers satisfy the RSGA with the same
 ladder spacing, the overall Hamiltonian naturally satisfies RSGA. A natural example
is Eq. (7) in the main text, which can be viewed as a ring formed by
connecting many identical AA dimers end-to-end: 
\begin{eqnarray}
H_{\text{ring}} &=&\sum_{j=1}^{N}[\omega a_{j}^{\dag }a_{j}+\Omega
s_{j}^{z}+g\left( a_{j}^{\dag }s_{j}^{-}+a_{j}s_{j}^{+}\right) +
Ua_{j}^{\dag }a_{j}s_{j}^{z}-\kappa \left( a_{j}^{\dag }a_{j+1}+\text{%
\textrm{H.c.}}\right) ],  \notag \\
&=&\sum_{j=1}^{N}H_{j,j+1}^{\mathrm{AA}},
\end{eqnarray}
where the expression for $H_{j,j+1}^{\mathrm{AA}}$ is given by: 
\begin{eqnarray}
H_{j,j+1}^{\mathrm{AA}}&=&h_{j}^{\text{a}}+h_{j+1}^{\text{a}}-\kappa\left(a_{j}^{%
\dagger}a_{j+1}+a_{j+1}^{\dagger}a_{j}\right),  \notag \\
h_{j}^{\text{a}} &=& \frac{\omega}{2}a_{j}^{\dagger }a_{j}+\frac{\Omega}{2}%
s_{j}^{z}+\frac{g}{2}\left( a_{j}^{\dagger }s_{j}^{-}+a_{j}s_{j}^{+}\right)+%
\frac{U}{2}s_{j}^{z}a_{j}^{\dagger}a_{j}.
\end{eqnarray}
According to Eq. (\ref{AA_Condition}), the condition for $H_{j,j+1}^{\mathrm{AA}}$ to
possess RSGA is: $U=\left(\Omega-\omega+2\kappa\pm\sqrt{\left(\Omega_{}-%
\omega+2\kappa\right)^2+2g^2}\right)$.

\subsection{Ground-state proof for the exact states under specific conditions}
\label{C}

In this section, we prove that, under the resonance conditions
established in Sec.~B, the constructed exact states are ground
states within their respective fixed-total-excitation sectors
when $\kappa>0$ and $U_l<0$ at every atom-doped site.
Throughout the proof, the JC couplings $g_l$ at atom-doped sites
are assumed to be real and nonzero.

We first recall a simple variational argument. If the same
normalized state $\ket{\Psi}$ is a ground state of every term
$H_\alpha$, then it is also a ground state of
$H=\sum_\alpha H_\alpha$. Indeed, for any normalized state
$\ket{\Phi}$,
\begin{equation}
\langle\Phi|H|\Phi\rangle
=\sum_\alpha\langle\Phi|H_\alpha|\Phi\rangle
\geq
\sum_\alpha\langle\Psi|H_\alpha|\Psi\rangle
=\langle\Psi|H|\Psi\rangle.
\label{GS_variational}
\end{equation}
Thus, $\ket{\Psi}$ attains the lowest energy of the full
Hamiltonian. This conclusion requires neither mutual
commutativity nor disjoint supports of the individual terms.
The same argument applies when all states and operators are
restricted to a common invariant subspace.

For an AA dimer, the Hamiltonian can be rewritten as
\begin{equation}
\begin{aligned}
H_{ij}^{\mathrm{AA}}
={}&
\left(h_i^{\mathrm a}-\kappa a_i^\dagger a_i\right)
+\left(h_j^{\mathrm a}-\kappa a_j^\dagger a_j\right)\\
&+\kappa(a_i-a_j)^\dagger(a_i-a_j).
\end{aligned}
\label{GS_dimer_decomposition}
\end{equation}
Here, the parameters entering $h_l^{\mathrm a}$ are the local
contributions assigned to a single dimer, as in Sec.~B.
The first two terms remain single-cavity JC--Stark Hamiltonians,
with the photon frequency shifted from $\omega_l$ to
$\omega_l-\kappa$, whereas the last term is positive semidefinite.
We retain the local generators $\eta_l$ and the common dimer
ladder spacing $\epsilon$ defined in Sec.~B.

We first determine the lowest energy of each shifted
single-cavity Hamiltonian. The resonance conditions imply
\begin{equation}
\epsilon=\omega_l-\kappa+\frac{U_l}{2},
\qquad
g_l^2=U_l(\epsilon-\Omega_l).
\label{GS_local_resonance}
\end{equation}
For a local excitation number $m\geq1$, the corresponding
subspace is spanned by
$\{\ket{g,m}_l,\ket{e,m-1}_l\}$.
The RSGA-protected state $\eta_l^m\ket{g,0}_l$ has energy
\begin{equation}
-\frac{\Omega_l}{2}+m\epsilon.
\label{GS_local_energy}
\end{equation}
Since this subspace is two dimensional, the other eigenvalue
is obtained by subtracting this exact eigenvalue from the trace
of the local Hamiltonian matrix. Its difference from the
RSGA-protected eigenvalue is
\begin{equation}
\Omega_l-\epsilon-mU_l
=-\frac{g_l^2}{U_l}-mU_l>0,
\label{GS_local_energy_difference}
\end{equation}
where the inequality follows from $U_l<0$.
Therefore, the RSGA-protected state is a ground state of the
shifted single-cavity Hamiltonian in the corresponding local
excitation sector. For $m=0$, the local vacuum is the only state
and has energy $-\Omega_l/2$.

Now consider the dimer sector with total excitation number $n$.
For any distribution of excitations between the two sites,
$m$ and $n-m$, the sum of the lowest energies of the two
shifted single-cavity terms is
\begin{equation}
\begin{aligned}
&\left(-\frac{\Omega_i}{2}+m\epsilon\right)
+\left(-\frac{\Omega_j}{2}+(n-m)\epsilon\right)\\
&\hspace{2cm}
=-\frac{\Omega_i+\Omega_j}{2}+n\epsilon.
\end{aligned}
\label{GS_dimer_local_bound}
\end{equation}
This lower bound is independent of the excitation distribution.
Consequently, any superposition of products of the corresponding
local ground states attains the same lower bound.
In particular, the dimer exact state
\begin{equation}
\ket{\psi_{ij}^{n}}
\propto
(\eta_i+\eta_j)^n\ket{\mathrm{Vac}}_{ij}
\label{GS_dimer_state}
\end{equation}
expands precisely into such products and is therefore a ground
state of the sum of the two shifted single-cavity terms within
the fixed-$n$ sector.

Furthermore,
\begin{equation}
[a_i-a_j,\eta_i+\eta_j]=0,
\qquad
(a_i-a_j)\ket{\mathrm{Vac}}_{ij}=0,
\end{equation}
which gives
\begin{equation}
(a_i-a_j)\ket{\psi_{ij}^{n}}=0.
\end{equation}
Thus, the same exact state also attains the zero-energy minimum
of the positive-semidefinite hopping term.
Applying the variational argument to these two contributions
establishes that $\ket{\psi_{ij}^{n}}$ is a ground state of
$H_{ij}^{\mathrm{AA}}$ within the fixed-$n$ sector, with energy
$-(\Omega_i+\Omega_j)/2+n\epsilon$.

The proof extends directly to the AC and CC dimers.
For an empty cavity, the resonance condition
$\varpi_l-\kappa=\epsilon$ reduces the shifted local term to
$\epsilon a_l^\dagger a_l$, whose energy in the $m$-photon sector
is $m\epsilon$.
Replacing an atom-doped site by an empty cavity therefore only
removes the corresponding atomic vacuum-energy contribution;
the remaining steps are unchanged.

For the arbitrary-lattice construction in Sec.~B, every dimer
has zero ladder spacing. Its entire family of exact states
therefore shares the same lowest energy.
Consider the collective exact state
\begin{equation}
\ket{\psi_{\mathrm E}^{n}}
\propto
\left(\sum_l\eta_l\right)^n\ket{\mathrm{Vac}}.
\label{GS_collective_state}
\end{equation}
Expanding this state according to the number of excitations
within any chosen dimer shows that its dimer component always
belongs to that dimer's ground-state family.
Hence the same collective state simultaneously minimizes
every dimer Hamiltonian, even when different dimers share sites.
The common-ground-state argument then gives
\begin{equation}
H_{\mathrm{res}}\ket{\psi_{\mathrm E}^{n}}
=E_{\mathrm{vac}}\ket{\psi_{\mathrm E}^{n}},
\qquad
E_{\mathrm{vac}}
=-\frac12\sum_l d_l\Omega_l,
\label{GS_lattice_energy}
\end{equation}
where, following the convention in the main text, $d_l$ counts
the dimers sharing site $l$, and $\Omega_l$ is understood to
vanish for empty cavities in this sum.
This proves the ground-state property of the negative-$U_l$
exact states for arbitrary lattice geometries and atom
distributions satisfying the stated resonance conditions.
The chemical-potential term introduced in the main text shifts
every energy in the fixed-$n$ sector by the same constant
$n\mu$. The constructed state therefore remains a ground state
within that sector, with energy $E_{\mathrm{vac}}+n\mu$.

Finally, consider the periodic ring, for which the dimer ladder
spacing need not vanish. Let $U_R<0$ denote the negative branch
of the ring resonance condition derived in Sec.~B.
Since each constituent dimer carries half of the onsite
parameters, its ladder spacing is
\begin{equation}
\epsilon=\frac{\omega}{2}-\kappa+\frac{U_R}{4}.
\label{GS_ring_dimer_spacing}
\end{equation}
Each site belongs to two dimers, so the sum of the excitation
numbers associated with all dimers is $2n$ for every
configuration with total excitation number $n$.
Summing the dimer energy bounds therefore yields, for any
normalized state $\ket{\Phi}$ in this sector,
\begin{equation}
\begin{aligned}
\langle\Phi|H_{\mathrm{ring}}|\Phi\rangle
&\geq-\frac{N\Omega}{2}+2n\epsilon\\
&=-\frac{N\Omega}{2}
+n\left(\omega-2\kappa+\frac{U_R}{2}\right).
\end{aligned}
\label{GS_ring_bound}
\end{equation}
The exact state constructed for the ring has precisely the
energy on the right-hand side and thus saturates this lower
bound. Consequently, at $U=U_R<0$, it is a ground state in each
fixed-total-excitation sector, without requiring the
ground-state energies of different sectors to coincide.

\subsection{Polariton condensates exhibit ODLRO}

\label{D} In this section, we compute the normalization
factor and the photon and atomic correlation functions
for the exact ground states of the uniformly atom-doped
ring, and establish their ODLRO in the thermodynamic limit.
We denote the unnormalized state as
\begin{equation}
\left\vert \tilde{\psi}_{\text{\textrm{G}}}^{n}\right\rangle =\left( \eta
\right) ^{n}\left\vert \mathrm{Vac}\right\rangle ,
\end{equation}%
where $\eta =\sum_{j=1}^{N}\left(ga_{j}^{\dagger }+U_{\mathrm{R}}s_{j}^{+}\right),\left\vert \mathrm{Vac}\right\rangle =\prod_{j=1}^{N}\left\vert 
\mathrm{Vac}_{j}\right\rangle ,\left\vert \mathrm{Vac}_{j}\right\rangle
=\left\vert g,0\right\rangle _{j}$. At $U=U_R$, the constructed state is a ground state of $H_{\text{ring}}$ within the sector with total excitation number $n$, where $U_R=\Omega-\omega+2\kappa-\sqrt{(\Omega-\omega+2\kappa)^2+2g^2}$. Using a generating-function
representation, the state can be equivalently written as%
\begin{equation}
\left\vert \tilde{\psi}_{\text{\textrm{G}}}^{n}\right\rangle =n!\left[ z^{n}%
\right] \exp \left( z\eta \right) \left\vert \mathrm{Vac}\right\rangle .
\end{equation}%
Here $\left[ z^{n}\right] \exp \left( z\eta \right) $ denotes the
coefficient of$\ z^{n}$ in the Taylor expansion of $\exp \left( z\eta
\right) $. The normalization factor is defined as 
\begin{eqnarray}
\Lambda _{n} &=&\left\langle \tilde{\psi}_{\text{\textrm{G}}}^{n}|\tilde{\psi%
}_{\text{\textrm{G}}}^{n}\right\rangle , \\
&=&\left( n!\right) ^{2}\left[ w^{n}z^{n}\right] \left\langle \mathrm{Vac}%
\right\vert e^{w\eta ^{\dagger }}e^{z\eta }\left\vert \mathrm{Vac}%
\right\rangle .
\end{eqnarray}%
Since operators on different sites commute, and the bosonic and spin
operators on the same site also commute, the vacuum expectation value
factorizes into a product of independent single-site contributions:%
\begin{equation}
\left\langle \mathrm{Vac}\right\vert e^{w\eta ^{\dagger }}e^{z\eta
}\left\vert \mathrm{Vac}\right\rangle =\prod_{j=1}^{N}\left\langle \mathrm{%
Vac}_{j}\right\vert e^{w\eta _{j}^{\dagger }}e^{z\eta _{j}}\left\vert 
\mathrm{Vac}_{j}\right\rangle.
\end{equation}%
The single-site vacuum expectation value evaluates to%
\begin{equation}
\left\langle \mathrm{Vac}_{j}\right\vert e^{w\eta _{j}^{\dagger }}e^{z\eta
_{j}}\left\vert \mathrm{Vac}_{j}\right\rangle =e^{wzg^{2}}\left( 1+wzU_{%
\mathrm{R}}^{2}\right) .
\end{equation}%
Therefore, the vacuum expectation value for the collective operators becomes%
\begin{equation}
\left\langle \mathrm{Vac}\right\vert e^{w\eta ^{\dag }}e^{z\eta }\left\vert 
\mathrm{Vac}\right\rangle =e^{Nwzg^{2}}\left( 1+wzU_{\mathrm{R}}^{2}\right)
^{N}.
\end{equation}%
Consequently, the normalization factor is given by%
\begin{equation}
\Lambda _{n}=\left( n!\right) ^{2}\left[ w^{n}z^{n}\right]
e^{Nwzg^{2}}\left( 1+wzU_{\mathrm{R}}^{2}\right) ^{N}.
\end{equation}%
Using the Cauchy product of series expansions, we obtain%
\begin{equation}
e^{Nwzg^{2}}\left( 1+wzU_{\mathrm{R}}^{2}\right) ^{N}=\sum_{n=0}^{\infty }%
\left[ \sum_{s=0}^{\min \left( n,N\right) }C_{N}^{s}U_{\mathrm{R}}^{2s}\frac{%
\left( Ng^{2}\right) ^{n-s}}{\left( n-s\right) !}\right] w^{n}z^{n},
\end{equation}%
Hence, the coefficient of$\ w^{n}z^{n}$ is%
\begin{equation}
\sum_{s=0}^{\min \left( n,N\right) }C_{N}^{s}U_{\mathrm{R}}^{2s}\frac{\left(
Ng^{2}\right) ^{n-s}}{\left( n-s\right) !}.
\end{equation}%
Therefore, the normalization factor is%
\begin{equation}
\Lambda _{n}=\left( n!\right) ^{2}\sum_{s=0}^{\min \left( n,N\right)
}C_{N}^{s}U_{\mathrm{R}}^{2s}\frac{\left( Ng^{2}\right) ^{n-s}}{\left(
n-s\right) !}.
\end{equation}%
Defining%
\begin{equation}
D_{n}^{\left( N\right) }=\sum_{s=0}^{\min \left( n,N\right) }C_{N}^{s}U_{%
\mathrm{R}}^{2s}\frac{\left( Ng^{2}\right) ^{n-s}}{\left( n-s\right) !},
\end{equation}%
we have$\ \Lambda _{n}=\left( n!\right) ^{2}D_{n}^{\left( N\right) }$, and
the normalized state reads $\left\vert \psi _{\text{\textrm{G}}%
}^{n}\right\rangle =\frac{1}{\sqrt{\Lambda _{n}}}\left( \eta \right)
^{n}\left\vert \mathrm{Vac}\right\rangle $.

The bosonic correlation function can be treated in the same way. For the
normalized state, it can be written as%
\begin{equation}
\left\langle \psi _{\text{\textrm{G}}}^{n}\right\vert a_{j}^{\dag
}a_{j+r}\left\vert \psi _{\text{\textrm{G}}}^{n}\right\rangle =\left\langle 
\tilde{\psi}_{\text{\textrm{G}}}^{n}\right\vert a_{j}^{\dag
}a_{j+r}\left\vert \tilde{\psi}_{\text{\textrm{G}}}^{n}\right\rangle
/\Lambda _{n}.
\end{equation}%
The numerator admits a generating-function representation,%
\begin{equation}
\left\langle \tilde{\psi}_{\text{\textrm{G}}}^{n}\right\vert a_{j}^{\dag
}a_{j+r}\left\vert \tilde{\psi}_{\text{\textrm{G}}}^{n}\right\rangle =\left(
n!\right) ^{2}\left[ w^{n}z^{n}\right] \left\langle \mathrm{Vac}\right\vert
e^{w\eta ^{\dag }}a_{j}^{\dag }a_{j+r}e^{z\eta }\left\vert \mathrm{Vac}%
\right\rangle .
\end{equation}%
The vacuum expectation value factorizes into single-site contributions, and
the insertion of $a_{j}^{\dag }a_{j+r}$ only modifies the factors at sites $j
$ and $j+r$. One finds 
\begin{eqnarray}
\left\langle \mathrm{Vac}\right\vert e^{w\eta ^{\dag }}a_{j}^{\dag
}a_{j+r}e^{z\eta }\left\vert \mathrm{Vac}\right\rangle  &=&\left[
wge^{g^{2}wz}\left( 1+U_{\mathrm{R}}^{2}wz\right) \right] \left[
zge^{g^{2}wz}\left( 1+U_{\mathrm{R}}^{2}wz\right) \right] \left[
e^{g^{2}wz}\left( 1+U_{\mathrm{R}}^{2}wz\right) \right] ^{N-2},  \notag \\
&=&g^{2}wz\sum_{n=0}^{\infty }\left[ \sum_{s=0}^{\min \left( n,N\right)
}C_{N}^{s}U_{\mathrm{R}}^{2s}\frac{\left( Ng^{2}\right) ^{n-s}}{\left(
n-s\right) !}\right] w^{n}z^{n}.
\end{eqnarray}%
Therefore, 
\begin{eqnarray}
\left\langle \tilde{\psi}_{\text{\textrm{G}}}^{n}\right\vert a_{j}^{\dag
}a_{j+r}\left\vert \tilde{\psi}_{\text{\textrm{G}}}^{n}\right\rangle 
&=&\left( n!\right) ^{2}g^{2}\left[ \sum_{s=0}^{\min \left( n-1,N\right)
}C_{N}^{s}U_{\mathrm{R}}^{2s}\frac{\left( Ng^{2}\right) ^{n-s-1}}{\left(
n-s-1\right) !}\right] ,  \notag \\
&=&\left( n!\right) ^{2}g^{2}D_{n-1}^{\left( N\right) }.
\end{eqnarray}%
Finally, the normalized correlator can be written compactly as
\begin{equation}
A_{r}=\left\langle \psi _{\text{\textrm{G}}}^{n}\right\vert a_{j}^{\dag
}a_{j+r}\left\vert \psi _{\text{\textrm{G}}}^{n}\right\rangle =g^{2}\frac{%
D_{n-1}^{\left( N\right) }}{D_{n}^{\left( N\right) }}.
\end{equation}
The expression for $A_{r}$ reduces to $A_{r}=n/N$  in the limit $U/g\rightarrow0$ , representing the correlation function of pure bosonic condensation.

The spin correlation function can be derived in an analogous way. For the
normalized state, it can be written as%
\begin{equation}
S_{r}=\left\langle \psi _{\text{\textrm{G}}}^{n}\right\vert
s_{j}^{+}s_{j+r}^{-}\left\vert \psi _{\text{\textrm{G}}}^{n}\right\rangle
=\left\langle \tilde{\psi}_{\text{\textrm{G}}}^{n}\right\vert
s_{j}^{+}s_{j+r}^{-}\left\vert \tilde{\psi}_{\text{\textrm{G}}%
}^{n}\right\rangle /\Lambda _{n}
\end{equation}%
The numerator admits a generating-function representation,%
\begin{equation}
\left\langle \tilde{\psi}_{\text{\textrm{G}}}^{n}\right\vert
s_{j}^{+}s_{j+r}^{-}\left\vert \tilde{\psi}_{\text{\textrm{G}}%
}^{n}\right\rangle =\left( n!\right) ^{2}\left[ w^{n}z^{n}\right] \left\langle 
\mathrm{Vac}\right\vert e^{w\eta ^{\dagger}}s_{j}^{+}s_{j+r}^{-}e^{z\eta
}\left\vert \mathrm{Vac}\right\rangle .
\end{equation}%
The vacuum expectation value again factorizes into single-site
contributions, and the insertion of $s _{j}^{+}s _{j+r}^{-}$ only
modifies the factors at sites $j$ and $j+r$. One finds 
\begin{eqnarray}
\left\langle \mathrm{Vac}\right\vert e^{w\eta ^{\dag
}}s_{j}^{+}s_{j+r}^{-}e^{z\eta }\left\vert \mathrm{Vac}\right\rangle 
&=&\left( e^{g^{2}wz}U_{\mathrm{R}}w\right) \left( e^{g^{2}wz}U_{\mathrm{R}%
}z\right) \left[ e^{g^{2}wz}\left( 1+U_{\mathrm{R}}^{2}wz\right) \right]
^{N-2},  \notag \\
&=&U_{\mathrm{R}}^{2}wz\sum_{n=0}^{\infty }\left[ \sum_{s=0}^{\min \left(
n,N-2\right) }C_{N-2}^{s}U_{\mathrm{R}}^{2s}\frac{\left( Ng^{2}\right) ^{n-s}%
}{\left( n-s\right) !}\right] w^{n}z^{n},
\end{eqnarray}%
To express the spin correlation function compactly, we introduce
\begin{equation}
\widetilde D_n^{(N)}
=
\sum_{s=0}^{\min(n,N-2)}
C_{N-2}^{s}U_{\mathrm R}^{2s}
\frac{(Ng^2)^{n-s}}{(n-s)!}.
\end{equation}
Unlike $D_n^{(N)}$, this coefficient contains only $N-2$
atomic factors, while the photonic contribution retains
the full factor $Ng^2$.

The unnormalized spin correlator therefore becomes
\begin{eqnarray}
\left\langle\tilde{\psi}_{\mathrm G}^{n}\right|
s_j^+s_{j+r}^-
\left|\tilde{\psi}_{\mathrm G}^{n}\right\rangle
&=&
(n!)^2U_{\mathrm R}^2
\left[
\sum_{s=0}^{\min(n-1,N-2)}
C_{N-2}^{s}U_{\mathrm R}^{2s}
\frac{(Ng^2)^{n-s-1}}{(n-s-1)!}
\right]
\notag\\
&=&
(n!)^2U_{\mathrm R}^2
\widetilde D_{n-1}^{(N)}.
\end{eqnarray}
Using $\Lambda_n=(n!)^2D_n^{(N)}$, the normalized correlator
for $n\geq1$ and $r=1,\ldots,N-1$ can be written as
\begin{equation}
S_r
=
U_{\mathrm R}^2
\frac{\widetilde D_{n-1}^{(N)}}{D_n^{(N)}}.
\end{equation}
For $0\leq n\leq N$, the pure-spin limit $g/U_R\to0$ gives
$S_r=n(N-n)/[N(N-1)]$ for distinct sites, as expected for the
symmetric atomic state with $n$ excitations.

We now establish ODLRO by deriving strictly positive lower bounds
for both correlation functions. We consider $N\geq2$ and
$n\geq1$, and take the thermodynamic limit at fixed filling
$\rho=n/N>0$, with real, nonzero $g$ and $U_R$ held fixed.
All coefficients $D_n^{(N)}$ and $\widetilde D_n^{(N)}$ with
nonnegative subscripts are then strictly positive. Coefficients
with negative subscripts are defined to vanish.

For the photon correlation, multiplying the defining sum for
$D_n^{(N)}$ by $n$ and splitting $n=(n-s)+s$ yields
\begin{equation}
\begin{aligned}
nD_n^{(N)}
={}&Ng^2D_{n-1}^{(N)}\\
&+NU_R^2
\sum_{s=0}^{\min(n-1,N-1)}
C_{N-1}^{s}U_R^{2s}
\frac{(Ng^2)^{n-1-s}}{(n-1-s)!}\\
\leq{}&N(g^2+U_R^2)D_{n-1}^{(N)}.
\end{aligned}
\label{ODLRO_D_bound}
\end{equation}
Here we used $sC_N^s=NC_{N-1}^{s-1}$ to obtain the second
term. The inequality follows from $C_{N-1}^{s}\leq C_N^s$
and the nonnegativity of every summand. Consequently,
\begin{equation}
A_r
=g^2\frac{D_{n-1}^{(N)}}{D_n^{(N)}}
\geq
\rho\frac{g^2}{g^2+U_R^2}
>0.
\label{ODLRO_A_bound}
\end{equation}

The same decomposition applied to $\widetilde D_n^{(N)}$
gives
\begin{equation}
Ng^2\widetilde D_{n-1}^{(N)}
\leq n\widetilde D_n^{(N)}
\leq
\left[Ng^2+(N-2)U_R^2\right]
\widetilde D_{n-1}^{(N)}.
\label{ODLRO_tildeD_bounds}
\end{equation}
The lower bound follows by retaining only the photonic
contribution, whereas the upper bound follows by bounding
the remaining atomic contribution. For $N=2$, the atomic
contribution is absent. Applying these inequalities at
successive indices gives
\begin{equation}
\begin{aligned}
\frac{\widetilde D_n^{(N)}}{\widetilde D_{n-1}^{(N)}}
&\leq
\frac{Ng^2+(N-2)U_R^2}{n}
\leq\frac{g^2+U_R^2}{\rho},\\
\frac{\widetilde D_{n-2}^{(N)}}{\widetilde D_{n-1}^{(N)}}
&\leq
\frac{n-1}{Ng^2}
\leq\frac{\rho}{g^2}.
\end{aligned}
\label{ODLRO_coefficient_ratios}
\end{equation}
For $n=1$, the second inequality holds directly by the
negative-subscript convention.

The two coefficient sequences also satisfy the exact identity
\begin{equation}
D_n^{(N)}
=
\widetilde D_n^{(N)}
+2U_R^2\widetilde D_{n-1}^{(N)}
+U_R^4\widetilde D_{n-2}^{(N)}.
\label{ODLRO_D_identity}
\end{equation}
This follows by expanding the two additional atomic factors
in their generating functions,
$(1+U_R^2x)^N=(1+U_R^2x)^{N-2}
(1+2U_R^2x+U_R^4x^2)$, where $x=wz$.
Combining this identity with
Eq.~(\ref{ODLRO_coefficient_ratios}) yields
\begin{equation}
\begin{aligned}
\frac{D_n^{(N)}}{\widetilde D_{n-1}^{(N)}}
&=
\frac{\widetilde D_n^{(N)}}{\widetilde D_{n-1}^{(N)}}
+2U_R^2
+U_R^4
\frac{\widetilde D_{n-2}^{(N)}}{\widetilde D_{n-1}^{(N)}}\\
&\leq
\frac{g^2+U_R^2}{\rho}
+2U_R^2+\frac{\rho U_R^4}{g^2}.
\end{aligned}
\label{ODLRO_ratio_bound}
\end{equation}
Since all quantities are positive, taking the reciprocal
provides the spin-correlation lower bound
\begin{equation}
\begin{aligned}
S_r
&=U_R^2
\frac{\widetilde D_{n-1}^{(N)}}{D_n^{(N)}}\\
&\geq
\frac{\rho g^2U_R^2}
{(g^2+\rho U_R^2)^2+g^2U_R^2}
>0.
\end{aligned}
\label{ODLRO_S_bound}
\end{equation}

Both lower bounds are independent of the system size and
the separation between distinct sites. Thus, at fixed
positive filling and fixed nonzero couplings, neither
correlation can vanish as the system size and the distance
between the two sites increase. Therefore, at fixed positive filling and fixed nonzero $g$
and $U_R$, both correlation functions remain finite and
nonzero at arbitrarily large separations in the thermodynamic
limit, establishing ODLRO in both the photon and atomic sectors.